\documentclass[twocolumn,amsmath,amssymb]{revtex4}

\usepackage{amsmath} %
\usepackage{graphicx}
\usepackage{amsfonts}
\usepackage{dcolumn}
\usepackage{bm}
\usepackage{color}

\begin{document}
\title{Disordered hyperuniform obstacles enhance sorting of dynamically chiral microswimmers}

\author{Jie Su}
\author{Huijun Jiang}
\thanks{E-mail: hjjiang3@ustc.edu.cn}
\author{Zhonghuai Hou}
\thanks{E-mail: hzhlj@ustc.edu.cn}
\affiliation{Department of Chemical Physics \& Hefei National Laboratory for Physical Sciences at Microscales, University of Science and Technology of China, Hefei, Anhui 230026, China}
\date{\today}

\begin{abstract}
Disordered hyperuniformity, a brand new type of arrangements with novel physical properties, provides various practical applications in extensive fields. To highlight the great potential of applying disordered hyperuniformity to active systems, a practical example is reported here by an optimal sorting of dynamically chiral microswimmers in disordered hyperuniform obstacle environments in comparison with regular or disordered ones. This optimal chirality sorting stems from a competition between advantageous microswimmer-obstacle collisions and disadvantageous trapping of microswimmers by obstacles. Based on this mechanism, optimal chirality sorting is also realized by tuning other parameters including the number density of obstacles, the strength of driven force and the noise intensity. Our findings may open a new perspective on both theoretical and experimental investigations for further applications of disordered hyperuniformity in active systems.
\end{abstract}

\maketitle
Hyperuniform point pattern performs a property that the variance of the point number contained within a regularly shaped observation window grows more slowly than its volume\cite{dh8,dh11}, differing from disordered patterns. Equivalently, a hyperuniform point pattern possesses a structure factor dependent on the small-wavenumber in a power-law form, i.e., $S(\mathbf{k})\sim |\mathbf{k}|^\alpha$ for $|\mathbf{k}|\rightarrow0$. Disordered hyperuniform (DH) patterns are states lying between crystals (regular patterns) and liquids (disordered patterns), which suppress large-scale density fluctuations like crystals, and lack any conventional long-range order like liquids or glasses\cite{dh26}. The positive exponent $\alpha$ represents a rough distinction between regular and DH patterns, i.e., the point pattern is DH for limited $\alpha$, but becomes stealthy\cite{dh12,dh13,dh14}, even turns into regular point pattern\cite{dh15} when $\alpha\rightarrow\infty$. In recent decades, DH structures have been widely discovered in many natural systems, including matter distribution in the Universe\cite{dh1,dh2}, avian cone photoreceptors\cite{dh3,dh4}, the ground state of liquid $^4$He\cite{dh5,dh6,dh7}, some aperiodic tilings\cite{dh1,dh8,dh9,dh10}, to list just a few. Besides, DH materials can also be conveniently synthesized in experiments by self-assembling of block-copolymer micelles\cite{dh22}, periodically driven emulsions\cite{dh23}, and many other methods\cite{dh24,dh25}. The distinct structure of DH patterns then provides various potential applications in extensive fields, such as surface-enhanced Raman spectroscopy\cite{dh16,dh17}, various photonic applications at the micro- and nanoscales\cite{dh18,dh19,dh20}, THz quantum cascade lasers\cite{dh21}, diffusion and conduction transport through DH media\cite{dh28,dh27}, etc.

As one of the hottest research topics in recent years across physical, chemical, materials and biological sciences, active systems have the ability to take in and dissipate energy from the environments so as to drive themselves far from equilibrium\cite{active1}. Active systems exhibit novel behaviors in comparison with their passive counterparts, such as emergence of dynamic chirality\cite{cap1,cap2,cap3,cap4,cap5}, active-induced phase separation\cite{active-ps,active-ps2}, collective vortex\cite{active2,active3}, polar swarms\cite{active7}, etc\cite{active4,active5,active6}. It is also reported that active particles show rather different behaviors in complex environments from passive ones\cite{active10,active12,active13,active14}. It is then a great opportunity to introduce the concept of disordered hyperuniformity to active systems, where new exciting behaviors may emerge.

In this paper, we report a practical example to highlight the potential of applying DH structures to active systems by a DH-pattern-enhanced sorting of dynamically chiral microswimmers (DCMSs). DCMSs taking a circular motion\cite{cap1,cap2,cap3,cap4,cap5} provides a routine for sorting of the microswimmers themselves or other particles coupling to them when they are driven through an obstacle environment\cite{cap6,cap9,cap10,cap11}, even in the situation where particle separation can be hardly achieved by mechanical means\cite{cap8}. Here, the sorting efficiencies are compared through a series of two-dimensional obstacle environments designed with patterns of regular, DH, or disordered types. Remarkably, an optimal sorting of DCMSs is observed in DH obstacles. Detailed analysis reveals that, as the pattern changes from regular to DH then disordered, on the one hand, the collision probability between DCMSs and obstacles increases, facilitating sorting of DCMSs, on the other hand, more and more DCMSs are found to be trapped by the obstacles, which hinders the efficiency of the chirality separation. The observed optimal chirality sorting in DH obstacles is then a consequence of the competition between these two effects. Based on this mechanism, the influences of the number density of obstacles and the strength of driven force are also investigated by further intensive simulations. Furthermore, it is found that there is also an optimal chirality sorting induced by noise, i.e., before the noise intensity increases to be large enough to dominate the dynamics and break the chirality sorting, noise can not only amplify the collision probability between microswimmers and obstacles, but also help those captured swimmers jump out of obstacle traps.

\begin{figure}
\includegraphics{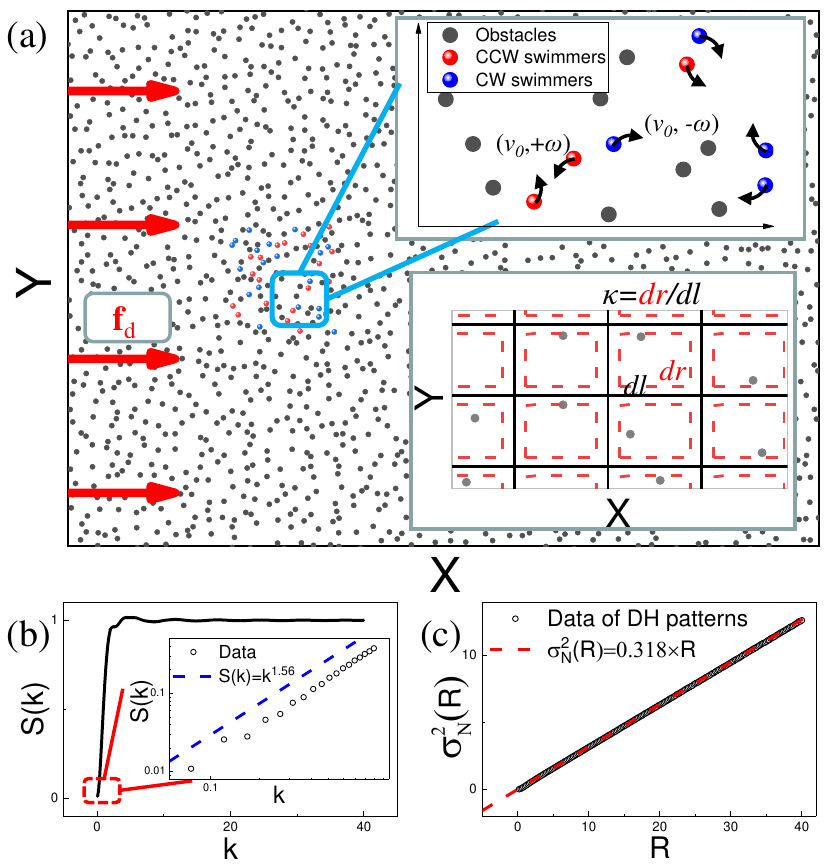}
\caption{\label{fig:model} Schematic of the system. (a) DCMSs (colored spheres) are driven through obstacle environments (gray points) by a static force (red arrows). The top inset is the zoom-in of the blue box with counterclockwise (CCW, red spheres) and clockwise (CW, blue spheres) microswimmers. The bottom inset shows the method to generate different patterns. (b) The structure factor $S(k)$ as a function of the wavenumber $k$ for generated DH obstacles. The inset is the zoom-in near $k\rightarrow0$. (c) The variance of point number $\sigma_N^2(R)$ as a function of the large window radius $R$.}
\end{figure}

The detailed setup of the system is shown in Fig.~\ref{fig:model}(a). A mixture of counterclockwise(CCW) and clockwise(CW) microswimmers are driven through an obstacle environment by a static force $\mathbf{f}_d=f_d\mathbf{e}_X$ parallelled to the X-axis. The CCW and CW swimmers have same radius $R_p$ and self-propelled rate $v_0$ except that the angular speed is $\omega$ for CCW swimmers and $-\omega$ for CW ones as shown in the top inset. The motion of each DCMS is described by the following overdamped Langevin equations

\begin{equation}
\frac{d\mathbf{r}_i}{dt}=v_0\mathbf{n}_i+\frac{1}{\gamma}[\mathbf{f}_d+\sum_{j,j\neq i}\mathbf{F}_{ij}^p(\mathbf{r}_{ij})+\sum_{m}\mathbf{F}_{im}^o(\mathbf{r}_{im})] +\mathbf{\xi}_i.\label{eq:translation}
\end{equation}
\begin{equation}
\frac{d\phi_i}{dt}=\omega_i+\mathbf{\zeta}_i.\label{eq:rotation}
\end{equation}

\noindent Herein, $\gamma$ is the friction coefficient, $\mathbf{r}_{ij(m)}=\mathbf{r}_i-\mathbf{r}_{j(m)}$, and $\mathbf{n}_i=(\cos(\phi_i),\sin(\phi_i))$ is the propulsion direction of swimmer i with $\phi_i$ the angle of $\mathbf{n}_i$. The interaction between DCMSs and that between DCMSs and obstacles are both of the linear spring form with the stiffness constants $k_p$ and $k_o$\cite{cap10,cap11} respectively, hence the force between the $i$th and $j$th swimmers is given by $\mathbf{F}_{ij}^p(\mathbf{r}_{ij})=k_p(2R_p-r_{ij})\Theta(2R_p-r_{ij})\mathbf{e}_{ij}$, and the force between the $i$th swimmer and the $m$th obstacle is $\mathbf{F}_{im}^o(\mathbf{r}_{im})=k_o(R_p+R_o-r_{im})\Theta(R_p+R_o-r_{im})\mathbf{e}_{im}$, with $r_{ij(im)}=|\mathbf{r}_{ij(im)}|$, $\mathbf{e}_{ij(im)}=\mathbf{r}_{ij(im)}/r_{ij(im)}$, $R_o$ the radius of the obstacle, and $\Theta(x)$ the Heaviside step function of $x$. $\mathbf{\xi}_i$ denotes the random force satisfying the fluctuation-dissipation relation $\langle\mathbf{\xi}_{\mu i}(t)\mathbf{\xi}_{\nu j}(t')\rangle=2D_t\delta(t-t')\delta_{\mu\nu}\delta_{ij}$, where the subscript $\mu$($\nu$) denotes the component along the X(Y)-axis, and $D_t$ is the translational diffusion coefficient. The last term $\zeta_i$ in Eqs. (\ref{eq:rotation}) is the rotational fluctuation satisfying $\langle\zeta_i(t)\zeta_j(t')\rangle=2D_o\delta(t-t')\delta_{ij}$ with $D_o$ the rotational diffusion coefficient.

To generate a two-dimensional DH array with number density $\rho_o$, we use a simple method inspired by Gabrielli et al.\cite{dh1} as follows: (Step 1)Cut the total surface into squares with same size $dl=\rho_o^{-1/2}$; (Step 2)Place a point randomly inside each square. An example of DH pattern consisting of $N_o=2500$ points generated by this method is plotted by gray points in Fig.~ \ref{fig:model}(a). The structure factor $S(k)$ averaged over $10^6$ ensembles is shown in Fig.~\ref{fig:model}(b). Clearly, it is found that $\lim_{k\rightarrow0}S(k)\sim k^\alpha$ with an exponent $\alpha=1.56\pm0.03$ (the inset of Fig.~\ref{fig:model}(b)), indicating the DH characteristic of the generated pattern. Moreover, as shown in \ref{fig:model}(c), the number variance $\sigma_N^2(R)\equiv\langle N_{\Omega}^2\rangle-\langle N_{\Omega}\rangle^2$ inside a large window $\Omega$ with side $R$ depends linearly on $R$, further demonstrating that the pattern is of DH type. Note that, it is also convenient to generate regular or disordered pattern by rescaling the squares in Step 2, i.e., randomly place a point in a square of size $dr$ (enclosed by red dashed lines in the bottom inset of Fig.~\ref{fig:model}(a)) which could be larger or smaller than $dl$. Naturally, the parameter $\kappa=dr/dl$ can be used to characterize the degree of disorder of the patterns. With this construction, the generated pattern is regular for $\kappa=0$, DH for $\kappa$ around 1, and disordered for large enough $\kappa$.

In simulations, parameters are made dimensionless by using $\gamma$, $v_0$ and $\omega$ as the basic units, so that the basic units for time, length and energy are $\omega^{-1}$, $v_0\omega^{-1}$ and $v_0^{2}\omega^{-1}\gamma$. According to the experimental data for DCMSs\cite{cap4,eq3}, we fix $R_p=R_o=0.5$, $D_t=5.0\times10^{-3}$, and $D_o=1.0^{-2}$, if not otherwise stated. Other parameters are $k_p=k_o=200$, $f_d=0.2$, $\rho_o=0.0977$ and time step $dt=10^{-3}$. For consistence, all of the following results are obtained from a mixture of $2000$ CCW and $2000$ CW swimmers running for a long time $t_{final}=4\times10^4$. The initial positions of those DCMSs are randomly chosen in a rectangle ($200\times100$) near the origin.

\begin{figure}
\includegraphics{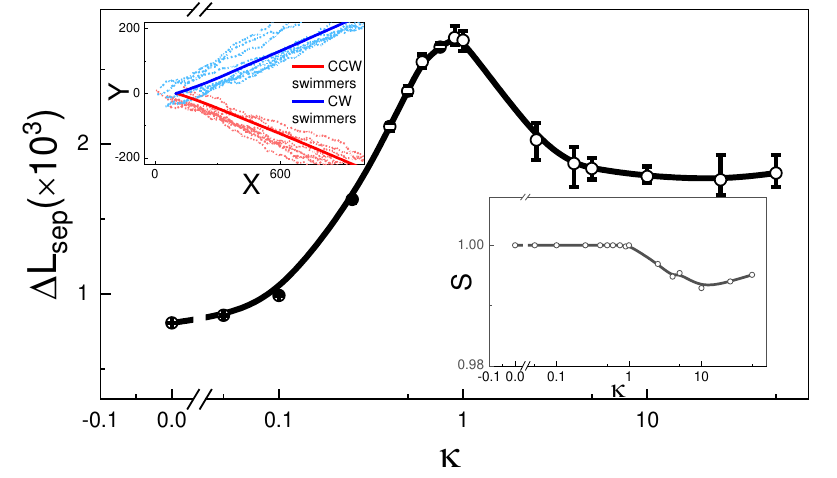}
\caption{\label{fig:result2} The difference between the centers of the CW and CCW groups in the Y-direction $\Delta L_{sep}$ as a function of the degree of disorder $\kappa$. The left inset is typical individual trajectories (light red and blue dashed lines) and average ones (dark red and blue solid lines) over all CCW (or CW) swimmers for $\kappa=1.0$ (DH patterns). The right inset shows the dependence of the sorting selectivity $S$ on $\kappa$.}
\end{figure}

First of all, we are interesting in how the degree of disorder of the obstacle affects the collective behaviors of DCMSs. Typical trajectories of individual swimmers as well as the averaged ones over all CCW (or CW) swimmers for $\kappa=1.0$ are plotted in the left inset of Fig.~\ref{fig:result2} (also can be seen in movie S1). It is observed that all the CCW swimmers move along the negative direction of Y-axis while all the CW swimmers move along the positive one, resulting a clear sorting of DCMSs.

In order to quantitatively measure the efficiency of chirality sorting, we introduce an order parameter
\begin{equation}
\Delta L_{sep}=\langle Y_{cw}(t_{final})\rangle-\langle Y_{ccw}(t_{final})\rangle,\label{eq:Lsep}
\end{equation}
where $Y_{cw(ccw)}$ denotes the Y-position of CW(CCW) swimmers, $\langle \cdot\rangle$ means averaging over all CW(CCW) swimmers, such that $\Delta L_{sep}$ shows the difference between centers of those two groups of DCMSs in Y-direction. Dependence of $\Delta L_{sep}$ on the degree of disorder $\kappa$ of the obstacle environment is shown in Fig.~\ref{fig:result2}. It can be seen that, $\Delta L_{sep}$ increases as $\kappa$ increases from 0 to 1, and then trails off when $\kappa$ passes through 1, i.e., obstacle patterns with $\kappa=1$ most favorably enhance $\Delta L_{sep}$ to the largest level. Notice that $\kappa=1$ corresponds to obstacles with the DH feature mentioned above, the observation thus indicates that the sorting of DCMSs in a DH environment is indeed more efficient than the regular or disordered ones.

To check whether a complete sorting of DCMSs can be obtained, we introduce another order parameter $S$ to describe the sorting selectivity,
\begin{equation}
S=\int_0^{\infty}P_{cw}(Y)dY-\int_0^{\infty}P_{ccw}(Y)dY,\label{eq:Lsep}
\end{equation}
where $P_{cw(ccw)}(Y)$ is the probability distribution of $Y_{cw(ccw)}$. If all CW swimmers move to $Y>0$($Y<0$) and all CCW swimmers move to $Y<0$($Y>0$), i.e., swimmers with different dynamic chirality can be separated completely, one has $S=+1$($S=-1$). The obtained $S$ as a function of $\kappa$ is presented in the right inset of Fig.~\ref{fig:result2}. $S\approx1$ for $\kappa\leq1$, and decreases to be smaller than $1$ for $\kappa>1$, indicating that complete chirality sorting can be achieved for hyperuniform patterns rather than disordered ones. Therefore, the obstacle with DH feature is best for sorting efficiency of DCMSs with $100\%$ sorting selectivity.

\begin{figure}
\includegraphics{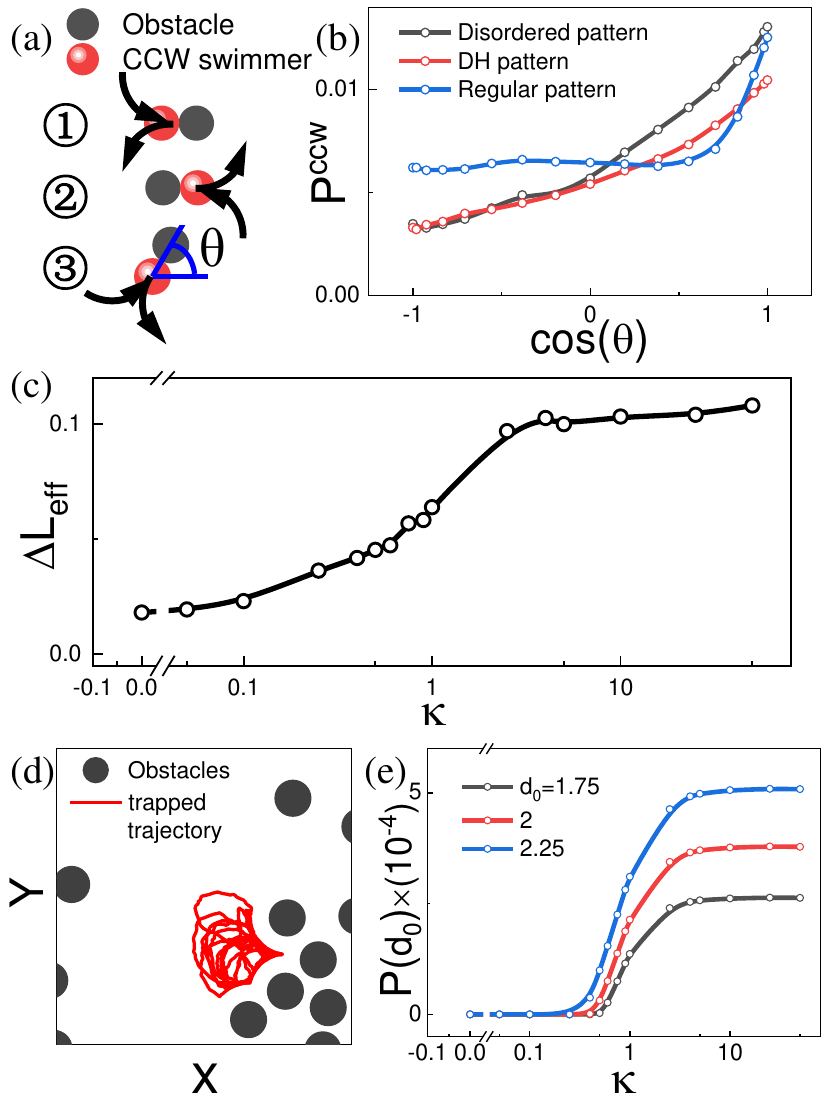}
\caption{\label{fig:result3} Underlying mechanism for the optimal sorting of DCMSs induced by DH obstacles. (a) Schematic of collisions for $\theta=0$, $\pi$ and a general value, respectively. (b) The possibility distribution function $P^{ccw}(\cos(\theta))$ for regular, DH or disordered patterns at $f_d=0.2$. (c) The effective separation length $\Delta L_{eff}$ in Y-direction as a function of $\kappa$. (d) Example of a CCW swimmer trapped in a disordered environment. (e) Possibility $P(d_0)$ of obstacle pairs whose distance is smaller than a given short length $d_0$ as functions of $\kappa$.}
\end{figure}

In order to figure out the underlying mechanism for the DH-pattern-induced optimal sorting of DCMSs, we first focus on the behaviors occurring in a single collision between DCMSs and obstacles. Taking CCW swimmers as an example, as presented in Fig.~\ref{fig:result3}(a), three cases are demonstrated for different collision angle $\theta$ which is defined as the angle between the vector pointing from the CCW swimmer to the obstacle and the X-axis. For $\theta=0$, the CCW swimmer would move along the negative direction of the Y-axis after collision, while it would move along the positive Y-direction for $\theta=\pi$. Generally, there is a shift $\Delta L_{ccw}$ in the Y-direction after a collision, which is greater than $0$ for $\theta\in(\pi/2,3\pi/2)$ ($\cos(\theta)<0$) and less than $0$ for $\theta\in(-\pi/2,\pi/2)$ ($\cos(\theta)>0$). The possibility distribution function $P^{ccw}(\cos(\theta))$ for regular, DH or disordered patterns at $f_d=0.2$ is then shown in Fig.~\ref{fig:result3}(b). It is clear that $P^{ccw}$ for $\cos(\theta)>0$ is larger than that for $\cos(\theta)<0$, indicating that CCW swimmers would move along negative direction of the Y-axis, keeping consistent with the trajectories of CCW swimmers in the inset of Fig.~\ref{fig:result2}. Moveover, $P^{ccw}$ of regular pattern is larger than the other two structures when $\cos(\theta)<0$, resulting in a smallest value of $\Delta L_{ccw}$ in regular obstacles. Meanwhile, $P^{ccw}$ of disordered pattern is larger than DH one when $\cos(\theta)>0$, leading to a largest value of $\Delta L_{ccw}$ in disordered obstacles. Correspondingly, all the results for CW swimmers are the same except that $\Delta L_{cw}>0$ for $\cos(\theta)>0$ and $\Delta L_{cw}<0$ for $\cos(\theta)<0$. To quantitatively characterize the effect of collision, we note that shifts after a collision satisfy $\Delta L_{ccw}=-2v_0\omega^{-1}\cos(\theta)$ for CCW swimmers and $\Delta L_{cw}=2v_0\omega^{-1}cos(\theta)$ for CW ones. For a small driven force such as $f_d=0.2$, the difference of collision-induced shifts in the Y-direction between CW and CCW swimmers can be measured approximately by $\Delta L_{eff}=\int_{-1}^{1}(\Delta L_{cw}P^{cw}-\Delta L_{ccw}P^{ccw})d\cos(\theta$). As plotted in Fig.~\ref{fig:result3}(c), $\Delta L_{eff}$ grows monotonically for increasing $\kappa$, indicating that if only independent collisions are taken into account, larger $\kappa$ would enhance the sorting efficiency of DCMSs.

\begin{figure}
\includegraphics{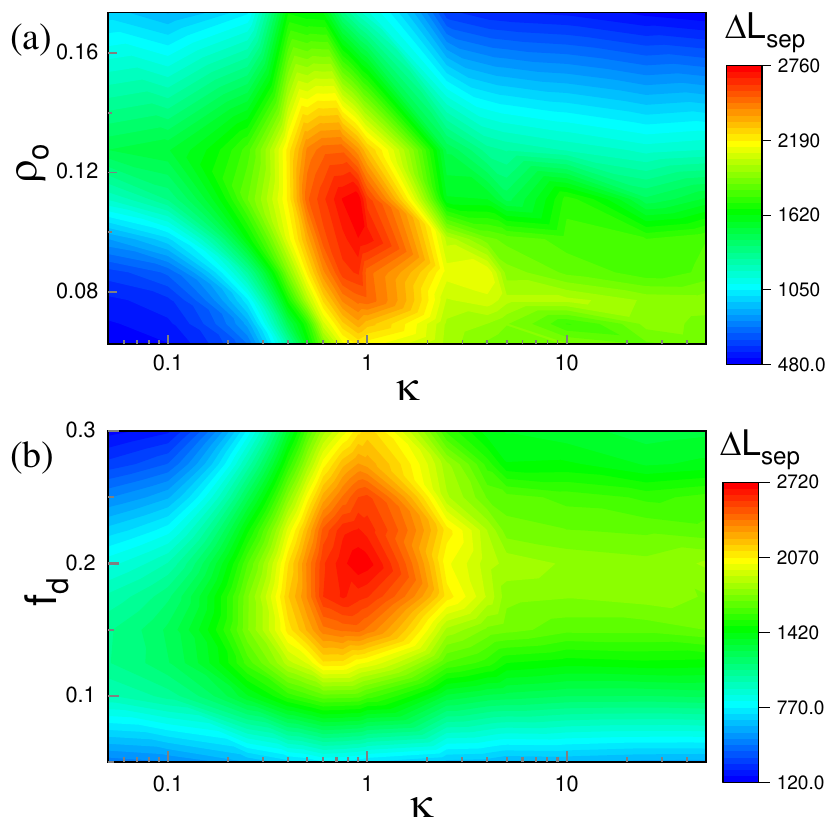}
\caption{\label{fig:result5} Sorting of DCMSs in environments for different obstacle densities and driven forces. The phase diagram for sorting of DCMSs in the $\kappa$-$\rho_o$ (a) and $\kappa$-$f_d$ (b) plane.}
\end{figure}

Nevertheless, besides of the per-collision effect, obstacles may also trap DCMSs by multiple collisions if several obstacles are near enough. As an example shown in Fig.~\ref{fig:result3}(d), within an obstacle environment with $\kappa=50$, a CCW swimmer is trapped in four neighbouring obstacles for a long time. The trapping effect can be more easily observed for $\kappa=50$(disordered patterns, movie S2) than $\kappa=1$(DH patterns, movie S1) or $\kappa=0$(regular patterns, movie S3). In order to depict how the structure of obstacle environment affects the frequency for traps occurring, we calculate the possibility $P(d_0)=2N_{d_0}/[N_o(N_o-1)]$ of obstacle pairs whose distance is smaller than a given short length $d_0$ for different $\kappa$ (Fig.~\ref{fig:result3}(e)). For all chosen $d_0$, $P(d_0)$ grows monotonically as $\kappa$ increases, indicating that as the environment becomes more and more disordered, DCMSs are more easily trapped, consequently hindering the efficiency of chirality sorting. Therefore, the competition between the advantageous per-collision effect and the disadvantageous multi-collision trapping effect results in an optimal sorting of DCMSs for $\kappa$ near $1$, i.e., the obstacle with DH feature.

To fully explore how parameters, including the degree of disorder $\kappa$, the number density of obstacles $\rho_o$ and the driven force $f_d$, affect the sorting of DCMSs, we have extensive simulations to obtain the phase diagrams in the $\kappa$-$\rho_o$ and $\kappa$-$f_d$ plane (Fig.~\ref{fig:result5}). Interestingly, peaks for $\Delta L_{sep}$ are observed in both of these two phase diagrams, indicating that the best sorting efficiency of DCMSs can be achieved by tuning these three parameters. The optimal chirality sorting for varying $\rho_o$ or $f_d$ can also be understood based on the aforementioned mechanism as follows. As $\rho_o$ or $f_d$ increases, although the advantageous per-collision effect between DCMSs and obstacles increases, the disadvantageous trapping effect of DCMSs due to multi-collision increases, either. Consequently, the competition of these two opposite effects finally leads to the optimal $\Delta L_{sep}$ dependent on $\rho_o$ or $f_d$.

\begin{figure}
\includegraphics{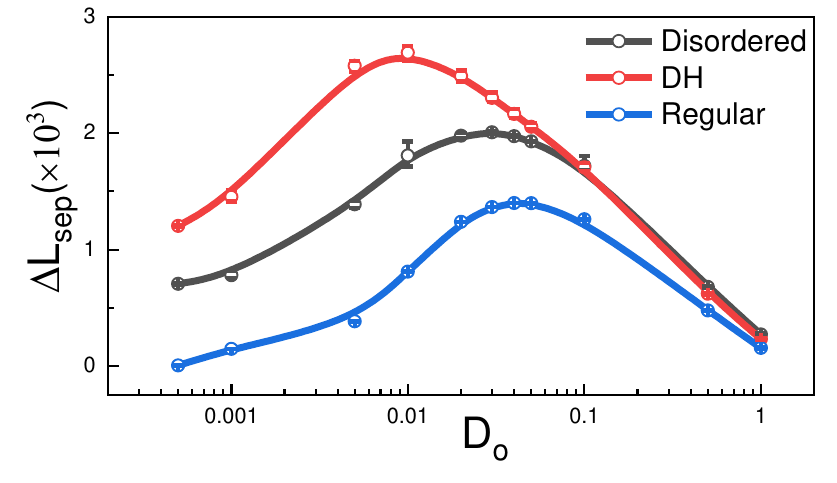}
\caption{\label{fig:result4} Dependence of chirality sorting on the noise intensity presented by $\Delta L_{sep}$ as functions of $D_o$ for different patterns.}
\end{figure}

Furthermore, according to the above understanding, one can also be predicted that noise would also be an important factor for sorting of DCMSs similar to $\kappa$. As the noise intensity increases, on the one hand, not only the probability of collisions increases, but also DCMSs would jump more easily out of obstacle traps, so that a more efficient sorting of DCMSs can be achieved, on the other hand, collisions becomes less and less effective, hindering the chirality separation. Therefore, based on the competition between these two effects, the chirality sorting efficiency should also achieve an optimal value for varying noise intensity. This prediction is then verified by further simulations with varied noise intensity $D_o$ and fixed $D_t/D_o=0.5$. Obtained dependence of $\Delta L_{sep}$ on $D_o$ for obstacles with disordered($\kappa=50$), DH($\kappa=1$) and regular($\kappa=0$) patterns is plotted in Fig.~\ref{fig:result4}. As expected, optimal $\Delta L_{sep}$ dependent on $D_o$ can be observed in all of the three patterns, demonstrating the constructive role noise plays in the sorting of DCMSs through obstacles.

In summary, we found that the sorting of DCMSs driven by a static force performs optimally through an obstacle environment of DH type in comparison with those of regular or disordered ones. This optimal chirality sorting results from the competition between obstacles-induced advantageous collisions and disadvantageous traps of DCMSs, both of which occur more frequently as the environment changes from regular to DH then disordered patterns. Based on the same competition of the two effects, it was predicted that, similar optimal chirality sorting should also be observed as the number density of obstacles, the driven force, or the noise intensity, which was demonstrated by further intensive simulations. The current study provides a convincing example to highlight the great potential of applying disordered hyperuniformity to active systems. Since systems designed with a DH pattern are experimentally available, and active systems are one of the hottest research areas across physical, chemical, materials as well as biological sciences, our findings may open a new perspective on both theoretical and experimental investigations for further applications of disordered hyperuniformity in active systems in the future.

This work is supported by MOST (2016YFA0400904, 2018YFA0208702), by NSFC (21833007, 21790350, 21673212, 21521001, 21473165), by the Fundamental Research Funds for the Central Universities (2340000074), and Anhui Initiative in Quantum Information Technologies (AHY090200).


\end{document}